# Effect of Strain disorder on the magnetic glassy state in $La_{5/8-y}Pr_yCa_{3/8}MnO_3$ (y = 0.45) thin films

Dileep K. Mishra[1], V. G. Sathe[1], R. Rawat[1], V. Ganesan[1], Ravi Kumar[2] and T. K. Sharma[2]

1. UGC- DAE Consortium for Scientific Research, University Campus, Khandwa Road Indore-452001

2. Semiconductor Laser Section, Raja Ramanna Centre for Advanced Technology, Indore-452013, India

*Abstract*

Present study reveals that the free energy landscape of the $La_{5/8-y}Pr_yCa_{3/8}MnO_3$ (LPCMO) system could be modified by elastic strain interaction in the epitaxial thin films. Epitaxial LPCMO thin films of various thicknesses are grown on $LaAlO_3$ substrate by pulsed laser deposition. With increasing thickness, by virtue of island growth morphology, strain disorder is invoked in thin films during strain relaxation process. The length-scale of phase separation is found to be highly correlated with strain disorder. Magneto-transport measurements demonstrate that coherent strain stabilizes charge ordered insulating phase while strain disorder stabilizes metallic phase. Resistivity under cooling and heating in unequal field (CHUF) protocol exhibits lower value of freezing temperature for strain disordered films compared to bulk system. Raman spectroscopy reveals that the charge ordered insulating and ferromagnetic metallic phases are structurally dissimilar and possess $P2_1/m$ and R-3C like symmetries respectively. Interfaces between two phases strongly influence low temperature glassy metastable state resulting in different phase separation states in the LPCMO thin films.

# 1. INTRODUCTION

In manganites, mesoscopic phase coexistence associated with first order phase transition (FOPT) is believed to be responsible for observed colossal magnetoresistance (CMR) [1]. In these systems, the electronically and magnetically contrasting charge ordered-antiferromagnetic insulator (CO-AFI) and ferromagnetic metallic (FMM) phases are found to coexist over a wide range of temperature and magnetic field [2,3,4]. LPCMO is one such highly studied system, where phase separation is observed at the micrometer length scale [5,6,7]. Many researchers believe it to be a result of quenched disorder due to variation in ionic size of the constituent elements [8,9,10]. On the other hand, our recent magnetic force microscopy studies gave direct visual evidence of fluid-like growth of the ferromagnetic metal FMM phase [7] that can not be solely explained by broadened first order phase transition due to quenched disorder. Recently, on epitaxial films of LPCMO grown on $SrTiO_3$ substrates it is shown that extrinsic disorder due to strain relaxation process invokes phase separation at large length scale that is sufficient to cross percolation threshold and cause a metal insulator transition[11]. In this study, it was shown that instead of quenched disorder, local strain fields play an important role and strong coupling between the electronic and elastic degrees of freedom results in multi-phase coexistence at micrometer length scale [11]. In this system and in manganites in general the crystal structure is intimately connected with the magnetic and electronic structure through Jahn-Teller distortion [12,13]. In fact there is a debate about the nature of metal insulator transition; with groups claiming it to be a Mott insulator [14] while other groups claiming it not to be a pure Mott insulator but rather have major contribution of structural distortion in the form of Jahn-Teller distortion. In recent years it is claimed that the charge localization is predominantly due to structural distortion [15]. Therefore, local structural disorder plays crucial role in framing the properties in manganites. Ahn *et al* [16] has shown that it is the structural aspect that primarily causes multiphase coexistence, commonly seen in martensites showing solid-solid phase transformation. The multiphase coexistence originates from lattice degrees of freedom rather than charge density [14].

Another puzzling aspect of doped manganites is their relaxation behavior; with some of the phase separated manganites displaying slow relaxation features that hide experimentally the real equilibrium thermodynamic state of the system [17,18]. Previously it is shown in LPCMO that the relaxation process is glass like and the ground state of this system is FMM [19]. Rather most of these intriguing properties like, large scale phase separation associated with first order transition, relaxation behavior at low temperature and response of this system toward external stimuli like, magnetic fields, laser photons, can be explained in the frame work of "magnetic glass" state [20, 21]. The micrometer scale phase separation can be explained by considering arrest of kinetics across the first order magnetic transition. The first order phase transition can be fully or partially arrested at low temperatures; the arrest occurs as one cools below a temperature $T_g$ called the magnetic glass transition temperature [22] which is concurrent with the classical liquid glass transition temperature when one cools the liquid at a higher rate so that one passes the transition temperature without crystallization.

The glass transition is a universal phenomenon which in its essence is independent of the interactions between the atoms or molecules of the liquid forming the glass. Here upon rapid cooling freezes a degree of freedom. Hence, it is necessary to state the degree of freedom that gets frozen during the magnetic glass transition in this system. P.A. Sharma et. al.[23] had proposed that structural interfaces comprise the degree of freedom that freezes at this unusual glass transition. It is observed that the time scale for the interface motion is in the microsecond range and is distinct from the time scale for relaxation of the relative FMM/CO-AFI volume fraction. However, the reason for freezing of this structural interface is not very clear. One of the possible explanations for freezing of interface boundaries can be elastic strain analogous with that observed in martensites. Here the elastic strain energy arising due to interfaces between two different crystal structures compete with the energy gain due to spin configuration (FM over AFI). Therefore it is important to study local structural changes and related interfaces that occur during CO-AFI to FMM phase transition and freezing of interface motion.

In this paper, we have shown that strain interactions are able to control length-scale of phase separation in LPCMO thin films. By employing Raman spectroscopy it is shown that the two FMM and CO-AFI coexisting phases are structurally dissimilar. The interface motion between the two structurally dissimilar states is getting affected due to elastic strain that can be modified in thin film configuration in the form of strain disorder by controlling strain relaxation process. The freezing of the interfacial motion degree of freedom resulting in arrest of kinetics is found to be strongly influenced by strain disorder. The devitrification of the arrested magnetic glassy phase (CO-AFI) has been studied using specially designed 'cooling and heating in unequal field' (CHUF) protocol [21,24].

## 2. EXPERIMENTAL

LPCMO thin films of various thicknesses have been deposited on commercially procured single crystal LaAlO$_3$ (LAO) substrate by Pulsed laser deposition (PLD) technique. Well characterized polycrystalline La$_{5/8-y}$Pr$_y$Ca$_{3/8}$MnO$_3$ (y = 0.45) is used as a target material during deposition. Lattice parameters of bulk target are (a = 5.435Å, b = 7.663Å and c = 5.4285Å in orthorhombic (Pnma) unit cell). Lattice constant of LAO substrate is (a$_s$) 3.79Å for pseudocubic unit cell. The films were prepared using a 248 nm KrF laser operating at 5 Hz frequency and 2 J cm$^{-2}$ energy density on the target. During deposition, the substrate temperature and oxygen pressure were 700°C and 300 mtorr respectively. All films were in-situ post annealed in 1 atm oxygen pressure for 30 min and then cooled to 150 °C in 60 min. Number of films of different thicknesses and post annealing conditions are prepared in order to realize different strain states in these films. All the films are well characterized by X-ray diffraction (XRD), Reciprocal space mapping (RSM) and epitaxy of the films is confirmed by φ-scan measurements. In order to study growth morphologies, all the films are characterized by atomic force microscopy (AFM) using digital instruments Nanoscope E in Contact Mode and the scan area is 1μm×1μm. Surface roughness, grain size and island height variations have been calculated by section analysis using Nanoscope 5.30 r1 software. Strong strain field inhomogeneities has been mainly considered as strain disorder and is being characterized in terms of island height variation and surface roughness reflected in AFM and XRD studies. Temperature dependent magneto-transport properties are studied by standard four probe resistivity methods in the quantum design cryostat and 8 Tesla superconducting magnet where magnetic field and current directions were kept parallel to each other. Detailed magneto-transport measurements were carried out by following various paths in H-T space. Raman spectroscopy has been carried out from 300K to 80K using LABRAM HR-800 spectrometer equipped with 488 nm and 633 nm excitation laser source, 1800 g/mm grating and a CCD detector giving an spectral resolution of ~1 cm$^{-1}$.

## 3. Results:

### 3.1: X-ray diffraction

Left panel of Figure 1 shows the normal (θ-2θ) x-ray diffraction results of the films of different thicknesses. It shows out-of-plane reflections corresponding to (101) orientation only besides the peaks from substrate indicating that all films are oriented along out of plan direction and single phase in nature. Observation of Keissig fringes in the XRD pattern of film F4 certifies defect free atomic level uniform film growth [11]. The out of plane reflection in all the XRD patterns are observed at lower 2θ values compared to substrate peak demonstrating that the films grow with in-plane compressive strain. The strain relaxation is clearly reflected in the XRD patterns of the films of higher thicknesses; at 100 nm the film (F4) is strained and with increasing thickness the film F3 (120nm) gets relaxed resulting in decrease in the out-of-plane lattice parameter. As the thickness is further increased, films F2 (175nm) and F1 (250nm) show two separate peaks. The peak occurring at lower 2θ values nearly matches with that for F4 film while 2θ position of the second peak (close to the substrate peak) matches exactly with the (202) peak position of the bulk compound in orthorhombic unit cell (shown as topmost curve in fig.1-a). Hence, peaks lying at lower 2θ and close to substrate peak are representing strained and relaxed phase respectively. In comparison to F2, for film F1 the intensity of relaxed phase is higher compared to strained phase indicating that relaxed phase grows with increasing thickness. In order to check the epitaxial nature of the films, ϕ-scan and reciprocal map (RSM) studies on these films has also been carried out. Representative RSM for F1 and F4 is shown in the right panel of the figure 1. RSM analysis further supports the inference that F4 is fully strained and F1 showing simultaneous presence of strained and relaxed phase. Horizontal and vertical width of contour representing strained phase is comparable in film F4 and F1 while contour representing relaxed phase in film F1 is extremely broad demonstrating that in-plane and vertical coherence length is very small for relaxed phase compared to strained phase. This establishes that near substrate regions of film is strained with larger coherence length while relaxed phase constrained within small islands.

### 3.2. Atomic force microscopy:

The growth morphology of the films is analyzed using AFM studies that are shown in figure 2. Upper panel of the figure reflects studies of thin film growth mechanism. Thin film growth mode has been optimized by increasing deposition time from 1 minute to 20 minutes. Careful surface analysis of these films shows that initial deposition (1-2minutes) results in uniform wetting of the substrate while further increase in deposition time to 5 min results in observation of highly dense sharp hillocks covering the surface of the film. On further increase in deposition time (10-20 minutes), agglomeration of these hillocks and formation of spherical shaped islands is observed confirming Stranski–Krastanov like thin film growth mode [25] shown as a schematic in the top right corner of the figure. Section analysis for the F1, F2, F3, and F4 films is given in the middle panel of the figure 2. Nearly 100 nm thick film is obtained after 30 minutes of deposition and named as F4 that shows highly uniform small spherical shaped islands. Surface roughness of this particular film is comparable to the substrate roughness (root-mean-square roughness ~5 Å) indicating highly uniform growth of the film as reflected in the observation of Keissig fringes in the XRD pattern for this film. Interestingly, on further increasing thickness (F3-120nm), the size of the islands is found to be comparable to that of the film F4. This indicates that on increasing thickness the island height gets increased that possibly invokes strain relaxation process as seen in the XRD studies. Further increment of thickness (120nm to 250nm) resulted in agglomeration of islands as evident in huge agglomeration observed in AFM images of the F2 and F1 films. Section analysis of the F2 and F1 films showed large grains with variable heights. A. Biswas et al[26] have shown that island growth morphology of the films lead to inhomogeneous distribution of strain, with strain relaxed top of the island and extremely strained periphery region of the island. Films F1 and F2 have huge islands and hence inhomogeneous strain field distribution at the periphery and top of the islands is anticipated in these films supported by simultaneous appearance of strained and relaxed phase in XRD. In the F1 and F2 film, each island have coexisting strained and relaxed phase resulting in strain disorder with in islands. Strain field distribution is schematized in lower panel of the figure. Uniform color represents uniform distribution of strain and color contrast represents strain field inhomogeneities within the islands.

### 3.3 Magneto-transport study:

**(a) Cooling and heating in equal field (CHEF):**

In order to study the insulator-metal transition ($T_{MI}$) phenomena and the associated PS, resistivity measurements as a function of temperature have been carried out on these films and shown in figure 3. The bulk LPCMO is insulating in 0T magnetic field, however, it shows $T_{MI}$ on the application of 2T field [11]. Presence of strain in F4 strengthens CO-AFI phase which favors insulating behavior in 0T and 2T magnetic field as reported before [11,19]. Application of 4T magnetic field destabilizes insulating phase and results in observation of $T_{MI}$ that is percolative in nature7. Resistivity behavior of film F3 is nearly similar to that for bulk compound11. On the other hand, the strain disordered films (F1, F2) show $T_{MI}$ in the 0T magnetic field contrary to bulk system. This $T_{MI}$ reflects transformation of CO-AFI to FMM state with orders of magnitude change in resistivity through first order phase transition. The F1 and F2 films show $T_{MI}$ (super cooling temperature T*) at 94K and 90K in cooling cycle receptively. In the warming cycle $T_{MI}$ (super heating temperature T**) of F1 and F2 are observed at 136 K and 132K respectively. Large hysteretic width (T** - T*) of 42K confirms first order nature of the transition. In 2T magnetic field, transition shifted to higher temperature and hysteresis width decreases. The low temperature behavior of these films where it shows an unusual upturn can be understood by considering arrest of kinetics across first order transition and magnetic glass state as reported before [7,19,11].

**(b) Cooling and heating in unequal field (CHUF):**

Detailed Magneto-transport measurements under the CHUF protocol [24, 21] has been used to elucidate the effect of strain disorder on the magnetic glassy behavior arising due to 'kinetic arrest' across CO-AFI to FMM transition. In present case, below ~$T_C$, CO-AFI phase is metastable magnetic glassy phase and devitrification of this glassy (CO-AFI) phase is observed during warming whenever warming field is higher than the cooling field[19,7]. All resistivity data sets for strain disordered films (F1, F2) shown in figure 4 (a,b) are collected during warming cycle using specially designed CHUF protocol in magneto-transport measurements traversing novel paths in H-T space (figure 4-c). Resistivity curves for which cooling field (0T, 0.5T, 1T) is lower than warming field (2T) showed sharp decrease in resistivity with increasing temperature (depicted by an arrow in figure 4-a) indicating devitrification of CO-AFI phase to FMM phase. This behavior is analogous to devitirification of structural glass during warming process. At a certain temperature mentioned as divitrification finish temperature ($T_{df}$) (table 1), the devitrification process of CO-AFI gets complete followed by FMM to CO-AFI first order phase transition at T**. The minima in the R-T curve is taken as $T_{df}$. In table1, $T_{df}$ (1T) and T* (2T) for strain disordered films (F1, F2) is compared with bulk like film (LPCMO/NGO)[19]. These results (table 1) revealed that in the presence of strain disorder divitrification is completed at lower temperature compared to bulk system indicating towards lower value of freezing temperatures for strain disordered films compared to bulk like film. The lower value of freezing temperature and higher value of T* helps in transformation of CO-AFI phase to FMM phase.

The transformed FMM phase is sufficient to cross percolation threshold even in the zero-field cooling cycle demonstrating metal-insulator transition in strain disordered films.

The results are best described with the phenomenological kinetic arrest and supercooling-superheating model [27,17] represented in figure 4(d). Supercooling and superheating temperatures could be varied with magnetic field in H-T space and represented by ($H^*$, $T^*$) and ($H^{**}$, $T^{**}$) lines. For disorder broadened first order phase transitions, super cooling and super heating lines are represented by ($H^*$, $T^*$) and ($H^{**}$, $T^{**}$) bands in H-T phase space as shown in figure 4(d) by blue and red bands respectively. In many reports anomalous thermomagnetic irreversibilities were observed well below the hysteretic regions of FOPT [28,29]. In our previous studied on LPCMO thin films, open loop behavior as well as virgin curve lying outside the envelope curve in isothermal R-H measurements were observed even at 5K[11,19] demonstrating that at low temperatures, kinetics of FOPT becomes critically slow (at experimental or laboratory time scales). Hence, during cooling CO-AFI phase gets kinetically arrested that did not transform completely to FMM ground state and form a metastable magnetic glass at low temperatures. The regions below which kinetics of transformation becomes arrested is represented by green colored ($H_K$, $T_K$) band in figure 4(d). In LPCMO system, ($H_K$, $T_K$) and ($H^*$, $T^*$) bands are anti-correlated to eachother[17]. This model displays that for strain disordered (F1, F2) films, ($H_K$, $T_K$) band lies within ($H^*$, $T^*$) band at zero field and hence during zero field cooling, regions crossing ($H^*$, $T^*$) before ($H_K$, $T_K$) would be transformed to FMM ground state and vice-versa resulted to arrested CO-AFI state. Because of higher overlap of ($H^*$, $T^*$) and ($H_K$, $T_K$) bands in strain disordered films, CO-AFI to FMM transformation crosses the percolation threshold even in zero-field cooling. It is worth noting here that sharp divitrification is observed for 0T cooled warming curves when compared to the divitrification for 0.5 T and 1T field cooled curves. This is because, in higher field cooling, less numbers of regions lying at higher temperature side of ($H_K$, $T_K$) band gets arrested compared to zero filed cooling. In case of cooling above a certain field ($H_2$) shown in figure 4(d), whole ($H^*$, $T^*$) band would be traversed before crossing the ($H_K$, $T_K$) band and system should be completely transformed to FMM phase.

### 3.4 Raman Spectroscopy:

The Raman measurements on manganites have been extensively carried out previously and the crystal structure responsible for even minute Raman features is available in literature[30,31,32,33, 34,35,36]. It is known that LPCMO belongs to rotationally distorted orthorhombic perovskites with space group (PNMA) that could be achieved from ideal cubic perovskite (Pm3m) by two consequent rotation of $MnO_6$ octahedra in two different directions, (010) and (101) of cubic perovskite structure. The crystal structure contains a network of corner sharing $MnO_6$ octahedra and rotated along ($a^-b^+a^-$) in glazer notation. Group theoretical phonon mode calculation for Pnma structure results in total 60 vibrational normal modes out of which 24 ($7A_g + 15B_{1g} + 17B_{2g} + 15B_{3g}$) are Raman active [36,37]. Figure 5 represent Raman spectra collected with 633 nm excitation laser line in the spectral range of (200-700 $cm^{-1}$) at room temperature. The Raman spectrum of all the films as well as the bulk reveals two broad peaks centered around 245 $cm^{-1}$ and 285 $cm^{-1}$ assigned as $A_g(4)$ and $B_{2g}(4)$ respectively following Illev *et. al*.[36] along with a broad hump lying between (375to 700 $cm^{-1}$). At room temperature due to extremely broad features of the Raman spectra it is difficult to assess the exact symmetry of the system, however, the spectra is in general agreement with that observed for many doped manganites with orthorhombic structure[31,34,36]. Appearance of two modes at 470 $cm^{-1}$ and 630 $cm^{-1}$ within hump are assigned as $A_g(3)$ and $B_{2g}(1)$ respectively. These modes correspond to in-phase asymmetric stretching (AS) and in-phase symmetric stretching (SS) of oxygen octahedral, and represent JT distortion present in the system [30-36]. The modes are sharp when long range cooperative JT distortion is present in the system and gets significantly broaden and lower in intensity for dynamical JT distortion along with huge background in the spectrum owing to electronic origin by presence of some delocalized carriers even at Room temperature [31].

In order to elucidate nontrivial resistivity behavior of the thin films, low temperature Raman study is performed on all the films across $T_{MI}$ using 488nm and 633nm excitation laser lines. Figures 6 (a-c) show temperature variation of Raman spectra for films F1 and F4 along with bulk compound using 488 nm excitation wavelength. The modes (AS) and (SS) induced due to JT distortion show concurrence with the insulator-metal transition upon cooling. As the temperature is lowered, there are nominal changes in the intensity of modes and at charge ordering temperature $T_{CO}$~ 210K [5], a sudden jump in the background is observed with intensity enhancement of JT distorted AS and SS modes. Intensity of these modes strengthens significantly in stained film (F4) and bulk compound along with appearance of new modes around 220, 400 and 420 $cm^{-1}$, evidencing structural modification upon charge-ordering. Further lowering of temperature down to 80K resulted in increase in the intensity of JT modes for insulating film F4 and bulk compound. On the other hand a sharp reduction in the intensity of JT modes accompanied with appearance of a new peak at 440 cm$^{-1}$ is observed for metallic film F1 below 120 K. In Figure 6 (d-e) evolution of intensity as well as line-width of the JT modes with lowering temperature is given. Importantly, line-width of the JT modes showed anomalous upturn below $T_{MI}$ for F1 while bulk and strain film shows normal decrease in line-width with lowering temperature. Anomalous temperature dependence of line-width below $T_{MI}$ is the signature of spin-lattice coupling as seen before in many manganite compounds.

Figures 7(a-d) demonstrate low temperature Raman spectra of all the films with 633 nm excitation. With lowering temperature, evolution of line-width as well as intensity of JT Raman modes show similar behavior as observed for 488 nm excitation. Additionally, for relaxed film (F3), these JT modes are slightly broader compared to strained film (F4) however, showed similar behavior with reducing temperature as that for film F4. Below 120K, for strain disordered films (F1 and F2), the line-width of JT modes increases dramatically along with steep reduction of intensity as observed with 488nm excitation. Furthermore, at 80K, the JT modes are extremely broad depicted by vertical arrows in figures 7(a-b) indicating dynamic nature of JT distortion. It is known that dynamic JT distortion is supported in rhombohedral structure [38,39] and static coherent long range JT distortion is supported in orthorhombic structure. This gives us evidence that F1, F2 films shows a structural rearrangement upon cooling across CO-AFI to FMM transition around 120 K with CO-AFI phase showing orthorhombic distortion while FMM phase showing rhombohedral structure. Another signature of this structural modification is seen in the occurrence of a peak at ~440 cm$^{-1}$ as depicted by an arrow in the figure for F1 and F2. Illiev *et. al.* [35] has clearly shown that this mode is present only in rhombohedral structure. A similar structural transition is seen in photo induced insulator to metal transition where this mode is taken as a signature of rhombohedral structure [38]. These films also showed anomalous changes in line position and width due to spin-lattice coupling [39].

Effect of excitation wavelength on the Raman spectra of all the films is clearly demonstrated in figure 7(e-h). For this the Raman spectra is collected at 80K using two wavelengths (633nm and 488nm) and compared. The first $A_g(4)$ mode falling at 260 cm$^{-1}$ showed stronger intensity for 488 nm when compared to 633 as expected. The 633 excitation wavelength showed predominantly stronger JT induced modes when compared to 488 nm excitation wavelength that is contrary to the expectation and is attributed to resonance effects [38,40]. The relative intensity of JT modes $I_{AS}/I_{SS}$ is high for 633 nm compared to that for 488 nm excitation. This shows that the JT band width is close to 2 ev as predicted before [38]. Due to resonance effects many of the features are seen more prominently in the Raman spectra collected using 633 nm excitation, for example, both the JT induced modes showed double peak features for insulating charge ordered film F4 that is absent in predominantly metallic film (F1, F2). These extra features are due to charge ordering that gets stabilized as a result of strain present in film F4. This is further supported by observation of extra phonon peaks depicted by arrows (in figure 7-h) as discussed before below charge ordering temperature. These peaks at 220 cm$^{-1}$, 400, 420 and 525 cm$^{-1}$ are prominent in strained film (F4) while absent in strain disordered films (F1, F2). Appearance of new phonon modes at low temperatures in strained film revealed that ordering of Mn$^{+3}$ and Mn$^{+4}$ modifies the crystal structure with formation of superlattice unit cell in charge ordered phase. Such observation are reported in previous studies mainly in half doped charge ordered system where CE-type charge ordering is well known. In the $La_{0.5}Ca_{0.5}MnO_3$ CE-type charge ordering occurs on lowering the temperature below $T_{CO}$ with superlattice unit cell of $P2_1/m$. Formation of superlattice unit cell resulted in folding of brillouin zone resulting in many additional lines in Raman spectra[31,33]. Additional lines are only prominent in strained film (F4) revealing that coherent strain stabilizes charge ordered monoclinic/orthorhombic phase. On the other hand, strain disordered films (F1 and F2) show structural changes from dominant orthorhombic phase to dominant rhombohedral phase with lowering of temperature across insulator to metal transition.

**4. Discussion:**

These results reveal that the free energy landscape of the system can be modified by elastic strain interaction in the LPCMO thin films. Different PS states have been achieved in the LPCMO films by strain relaxation process with increasing thickness. The thick films (F1 and F2) showed coexistence of strained and relaxed phases leading to very high degree of strain disorder in these films. Our Raman results on these films showed coexisting orthorhombic and rhombohedral structures at low temperatures. On the other hand the stained film (F4) showed monoclinic ($P2_1/m$) structure at low temperatures concurrent with CE type charge order state which is able to accommodate large strain values. The observation of coexisting crystal structures and gradual structural transformation around $T_{MI}$ in stain disordered films is similar to that observed in martensite like transformation. According to the model proposed by Ahn et.

al,[16] anisotropic lattice distortion results in insulating state while that with-out lattice distortion gives metallic phase and it is the structural aspect that causes the multiphase coexistence. Our study showed that strain disorder favors the FMM phase.

Our Raman study clearly revealed that coherent JT distortion suddenly gets suppressed below $T_{MI}$ along with observation of phonon mode typical of rhombohedral structure. The present Raman results clearly showed that across $T_{MI}$ the CO-AFI phase with orthorhombic/monoclinic structure coexist with FMM phase with rhombohedral structure. Simultaneous presence of two structurally dissimilar phases in PS regions is expected to create large interface energy in analogy to matensitic transition. In order to minimize interface energy, interfaces start to move and PS became dynamic in nature. In the dynamic PS state, on lowering the temperature equilibrium phase (FMM in this particular case) grows over the high temperature phase up to $T_g$. Below $T_g$, interface motion between two phases freezes[23]. Freezing of interface motion prevents the transformation of high temperature phase to equilibrium phase and metastable high temperature (orthorhombic CO-AFI) phase gets kinetically arrested forming "magnetic glass"[7,19]. The relaxation behavior of magnetic glassy phase in manganites can not be explained by considering it as a canonical ensemble; it can not be explained by considering it a frustrated system like spin glass or cluster glass with the direct visual evidence of presence of ferromagnetic regions at micron length scale[7] rather, it shows analogy with arrest of kinetics of liquid upon rapid cooling forming structural glass[18]. The degree of freedom of dynamic motion of molecules in liquid freezes in glassy state. Here in "magnetic glass" the degree of freedom of an individual atom is not freezing but the degree of freedom in the form of motion of regions that expands around transition temperature freezes resulting in a incomplete transition and the two phases co-exists to lowest measuring temperature (5 K). Sharma et al[23] attributed this to the freezing of structural interfaces and our Raman measurements provides direct evidence of two different structures existing in this compound and growth of one structure over the other as the system traverses across the CO-AFI to FMM transition. The changes in structure depend on the extrinsic disorder or inhomogeneous strain fields generated due to relaxation process of the films. In the F3, F4 films, the strain is homogeneous resulting in CO-AFI phase with concurrent orthorhombically distorted crystal structure and therefore, in these films the kinetically arrested CO-AFI phase remains dominating at lower temperature and system remains insulating till lowest possible measuring temperature. In F1 and F2 films, simultaneous presence of strained (P21/m) phase with relaxed (R-3C) phase generated greater number of structural interfaces in comparison to bulk system; as a result interface energy of the system will increase. For the growth of two phases these interfaces will acts as quenched disorder and PS at much larger length scales is favored. On lowering the temperature, rhombohedral structure is able to grow at much larger length scales and system crosses the percolation threshold and $T_{MI}$ is observed [11]. Due to larger number of interfaces in these films the freezing of interfacial motion or freezing of growth of FMM polarons is occurring at much lower temperature concomitant with glass temperature. This magnetic glass temperature is reflected in nearly constant resistivity value or a small upturn in resistivity below ~30 K.

## 5. Conclusions:

Our results clearly show that the elastic strain interactions among two structurally dissimilar phases play a major role in deciding the ratio of the two phase fractions present in the LPCMO thin films. In these thin films, presence of strain disorder modifies free energy landscape of the system that strongly influences the metastable glassy CO-AFI phase and favors equilibrium FMM phase. For the growth of two phases, structural interfaces will acts as quenched disorder and PS is favored at much larger length scales. The two phase fraction however, freezes below glass transition like the motion of a liquid freezes in a glassy state. Therefore, in magnetic glass state it can be said that the polaronic growth freezes below glass transition temperature.

**Acknowledgement:** The authors gratefully thank Dr. P Chaddah, for interest in this work and very fruitful discussions.

**References:**

[1] E. Dagotto New J. Phys. **7,** 67 (2005)
[2] S. Kumar and P. Majumdar, Phys. Rev. Lett. **92**, 126602 (2004).
[3] S. Yunoki, A. Moreo, and E. Dagotto, Phys. Rev. Lett. **81**, 5612 (1998).
[4] N. D. Mathur, P.B. Littlewood, Solid State Communications **119** 271-280 (2001).
[5] M. Uehara, S. Mori, C.H. Chen, And S-W Chang, Nature (London) **399**, 560 (1999).
[6] L. Zhang, C. Israel, A. Biswas, R. L. Greene, A. Lozanne, Science **805**, 298 (2002)
[7] R. Rawat, Pallavi Kushwaha, Dileep K. Mishra, and V. G. Sathe Phys. Rev. B **87** (2013) 064412.
[8] E Dagotto, Science **309**, 257, (2005).
[9] A. Moreo, M. Mayr, A. Feiguin, S. Yunoki, and E. Dagotto, Phys. Rev. Lett. **84**, 5568 (2000).
[10] Burgy, J., Moreo, A. & Dagotto, E. Phys. Rev. Lett. **92**, 097202 (2004).
[11] Dileep K Mishra, V.G. Sathe, R. Rawat, and V. Ganesan, J. Phys.: Cond. Matter **25** 175003 (2013).
[12] J. M. D. Coey, M. Viret & S. von Molnár, Advances in Physics, **48**, 167 (1999).
[13] A. J. Millis, Nature, **392**, 147 (1998).
[14] I. Loh, P. Adler, A. Grzechnik, K. Syassen, U. Schwarz, M. Hanfland, G. Kh. Rozenberg, P. Gorodetsky, and M. P. Pasternak, Phys. Rev. Lett. **87** (2001) 125501

[15] M. Baldini, V.V. Struzhkin, A. F. Goncharov, P. Postorino, and W. L. Mao, Phys. Rev. Lett. **106**, (2011) 066402

[16] K. H. Ahn, T. Lookman and A. R. Bishop, Nature (London) **428**, 401 (2004).

[17] Kranti Kumar, A. K. Pramanik, A. Banerjee, P. Chaddah, S. B. Roy, S. Park, C. L. Zhang, and S. W. Cheong Phys. Rev. B **73**, 184435 (2006).

[18] P. Chaddah, Kranti Kumar, and A. Banerjee, Phys. Rev. B **77**, 100402 (2008).

[19] V. G. Sathe, Anju Ahalawat, R. Rawat and P. Chaddah J. Phys. : Condens. Matter **22**, 176002 (2010).

[20] A . Lakhani, P. Kushwaha, R. Rawat, K. Kumar, A. Banerjee and P. Chaddah, J. Phys.: Condens. Matter **22**, 032101 (2010 ).

[21] A. Banerjee, K. Mukherjee, K. Kumar and P. Chaddah Phys.Rev. B **74**, 224445 (2006); A. Banerjee, A. K. Pramanik, K. Kumar and P. Chaddah, J. Phys.: Condens. Matter **18** L605 (2006).

[22] M. K. Chattopadhyay, S. B. Roy, and P. Chaddah, Phys. Rev. B **72**, 180401 (2005).

[23] P. A. Sharma, S. El-Khatib, I. Mihut, J. B. Betts, A. Migliori, S. B. Kim, S. Guha, and S.-W. Cheong, Phys. Rev. B **78**,134205 (2008).

[24] A Banerjee, Kranti Kumar and P Chaddah, J. Phys.: Cond. Matter **21**, 026002 (2009).

[25] Ivan N. Stranski and Lubomir Krastanow, Abhandlungen der Mathematisch-Naturwissenschaftlichen Klasse IIb. Akademie der Wissenschaften Wien, **146**, 797- 810 (1938).

[26] Amlan Biswas, M. Rajeswari, R. C. Srivastava, Y. H. Li, T. Venkatesan, R. L. Greene and A. J. Millis, Phys. Rev. B **61**, 9665 (2000); *ibid* Phys. Rev. B **63**, 184424 (2001).

[27] M. A. Manekar, S. Chaudhary, M. K. Chattopadhyay, K. J. Singh, S. B. Roy and P. Chaddah Phys. Rev. B **64**, 104416 (2001).

[28] R. Rawat, K. Mukherjee, K. Kumar, A. Banerjee and P. Chaddah J. Phys.: Condens. Matter **19**, 256211(2007); P. Kushwaha, A. Lakhani, R. Rawat, A. Banerjee, and P. Chaddah,Phys. Rev. B **79**, 132402 (2009).

[29] P. Kushwaha, R. Rawat and P. Chaddah J. Phys.: Condens.Matter **20**, 022204 (2008).

[30] M. V. Abrashev, J. Ba ckstrom, and L. Borjesson , Phys. Rev. B **64**, 144429.

[31] M. N. Iliev, M. V. Abrashev, V. N. Popov, and V. G. Hadjiev Phys. Rev. B **67**, 212301(2003).

[32] Martin-Carron, A. de Andres, M. J. Martinez-Lope, M. T. Casais, and J. A. Alonso, Phys. Rev. B **66**, 174303 (2002).

[33] M. Kim, H. Barath, S. L. Cooper, P. Abbamonte, E. Fradkin, M. Rübhausen, C. L. Zhang, and S.-W. Cheong Phys. Rev. B **77**, 134411 (2008)

[34] E. Liarokapis, Th. Leventouri, D. Lampakis, D. Palles, J. J. Neumeier and D. H. Goodwin, Phys. Rev. B **60**, 12758 (1999).

[35] M. N. Iliev and M. V. Abrashev, J. Raman Spectrosc.; **32**, 805–811 (2001).

[36] M. N. Iliev, M. V. Abrashev, H.-G. Lee, V. N. Popov, Y. Y. Sun, C. Thomsen, R. L. Meng and C. W. Chu, Phys. Rev. B 73, (2006) 064302

[37] M. N. Iliev, M. V. Abrashev, J. Laverdière, S. Jandl, M. M. Gospodinov, Y.-Q. Wang, and Y.-Y. Sun, Phys. Rev. B **57**, 2872 (1998).

[38] V G Sathe, R Rawat, Aditi Dubey, A V Narlikar and D Prabhakaran, J. Phys.: Condens. Matter **21**, 075603 (2009).

[39] Aditi Dubey, V. G. Sathe and R. Rawat, J. Appl. Phys. **104,** 113530 (2008), Aditi Dubey and V G Sathe, J. Phys.: Condens. Matter **19,** 346232 (2007).

[40] R Kruger, B Schulz, S Naler, R Rauer, D Budelmann, J Backstrom, K H Kim, S-W Cheong, V Perebeinos and M Rubhausen, Phys. Rev. Lett. **92,** 97203 (2004).

**Figure Captions:**

Figure 1: (a) θ-2θ X-ray diffraction results of LPCMO thin films of various thicknesses grown on LaAlO$_3$ substrate. Only (202) reflection of the films along with polycrystalline bulk is presented here. Thickness of the F1, F2, F3 and F4 films are 250 nm, 175 nm, 120 nm and 100 nm respectivly. (b) The reciprocal space maps of the F1 and F4 films along the (103) asymmetric diffraction spot. Dotted lines are position markers. $Q_x$(= hλ/2a) and $Q_z$(= lλ/2c) represent the in-plane and out-of-plane reciprocal lattice units (rlu).

Figure 2: Atomic force microscopy images of the of LPCMO thin films of various thicknesses grown on LaAlO$_3$ substrate. Upper panel represents surface morphology of initially deposited few layers as well as schematic for the Stranski–Krastanov like thin film growth mode. Middle panel shows island height variation across the diagonal of the 1µm×1µm scanned area of the F1, F2, F3 and F4 films (depicted by black lines). Lower panel presents schematic of strain distribution in LPCMO films.

Figure 3: Temperature dependent resistivity measurement performed during cooling and warming cycle in the presence of different magnetic fields for the strain disordered films (F1 and F2), relaxed film (F3) and strained film (F4).

Figure 4: (a), (b) Resistivity versus temperature in the warming cycle under the 2 T field from 5K to 240K after cooling in the labeled magnetic field. Panels (c): Representation of CHUF measurement protocol used for magneto-transport measurements with traversing novel paths in H-T space and panel (d) represents schematic diagram of kinetic arrest model in LPCMO thin films.

Figure 5: Room temperature Raman spectra of F1, F2, F3 and F4 films along with bulk compound in unpolarized geometry using He-Ne (632.8 nm) excitation source.

Figure 6: (a)-(c) The Low temperature Raman spectra of F1 and F4 films along with bulk compound collected from 300K to 80K using 488 nm excitation source. (d)-(e) Temperature dependence of the linewidth and intensity of the AS and SS modes for F1, F4 and bulk compound. Arrow depicts anomalous behavior of the SS Raman mode width for F1 film below $T_{MI}$.

Figure 7: (a)-(d) The Low temperature Raman spectra of all the four LPCMO films (F1, F2, F3 and F4) collected from 300K to 80K using 632.8nm excitation source. (e)-(g) Comparison of Raman spectra at 80K for all the four films collected using 632.8nm and 488nm excitation lines. Appearance of new peaks depicted by arrows revealing structural rearrangement below $T_{MI}$ and $T_{CO}$ for F1 and F4 films respectively.

**Figure 1:**

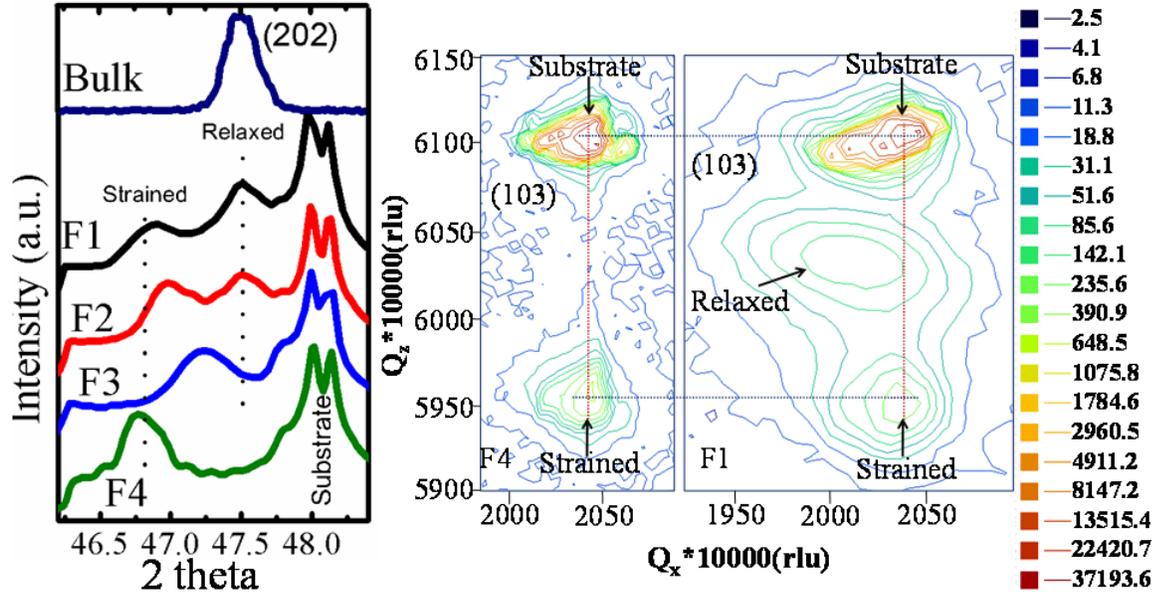

**Figure 1**: (a) θ-2θ X-ray diffraction results of LPCMO thin films of various thicknesses grown on LaAlO$_3$ substrate. Only (202) reflection of the films along with polycrystalline bulk is presented here. Thickness of the F1, F2, F3 and F4 films are 250 nm, 175 nm, 120 nm and 100 nm respectivly. (b) The reciprocal space maps of the F1 and F4 films along the (103) asymmetric diffraction spot. Dotted lines are position markers. $Q_x (= h\lambda/2a)$ and $Q_z (= l\lambda/2c)$ represent the in-plane and out-of-plane reciprocal lattice units (rlu).

**Figure 2:**

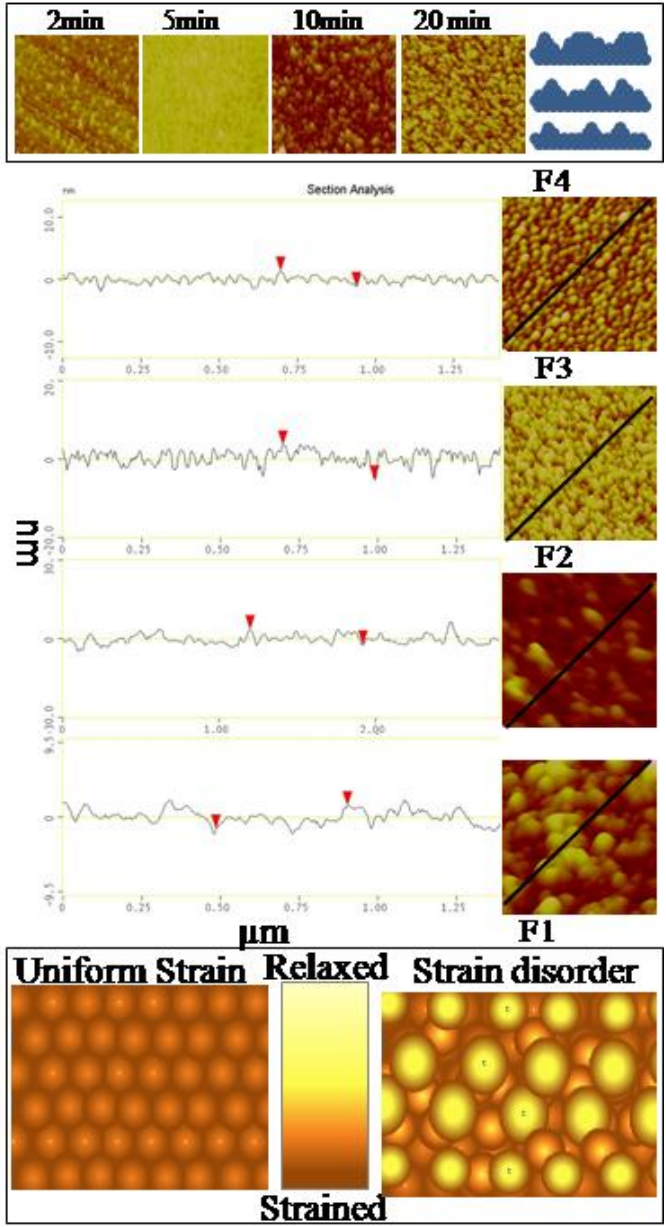

**Figure 2:** Atomic force microscopy images of the of LPCMO thin films of various thicknesses grown on LaAlO$_3$ substrate. Upper panel represents surface morphology of initially deposited few layers as well as schematic for the Stranski–Krastanov like thin film growth mode. Middle panel shows island height variation across the diagonal of the 1μm×1μm scanned area of the F1, F2, F3 and F4 films (depicted by black lines). Lower panel presents schematic of strain distribution in LPCMO films.

**Figure 3:**

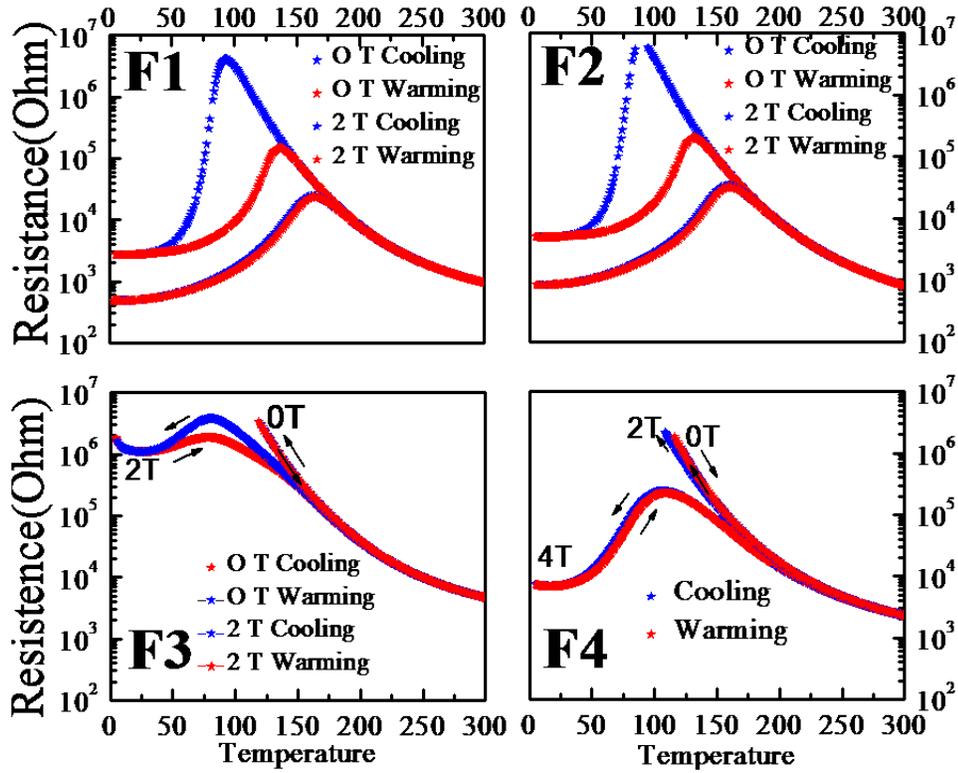

**Figure 3:** Temperature dependent resistivity measurement performed during cooling and warming cycle in the presence of different magnetic fields for the strain disordered films (F1 and F2), relaxed film (F3) and strained film (F4).

**Figure 4:**

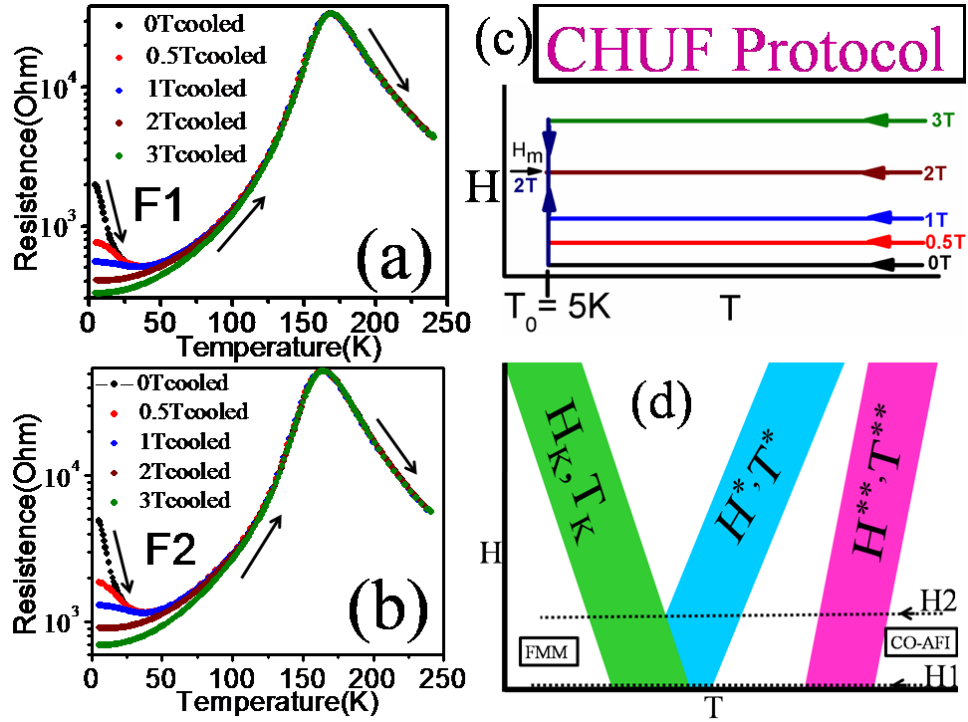

**Figure 4:** (a), (b) Resistivity versus temperature in the warming cycle under the 2 T field from 5K to 240K after cooling in the labeled magnetic field. Panels (c): Representation of CHUF measurement protocol used for magneto-transport measurements with traversing novel paths in H-T space and panel (d) represents schematic diagram of kinetic arrest model in LPCMO thin films.

**Figure 5:**

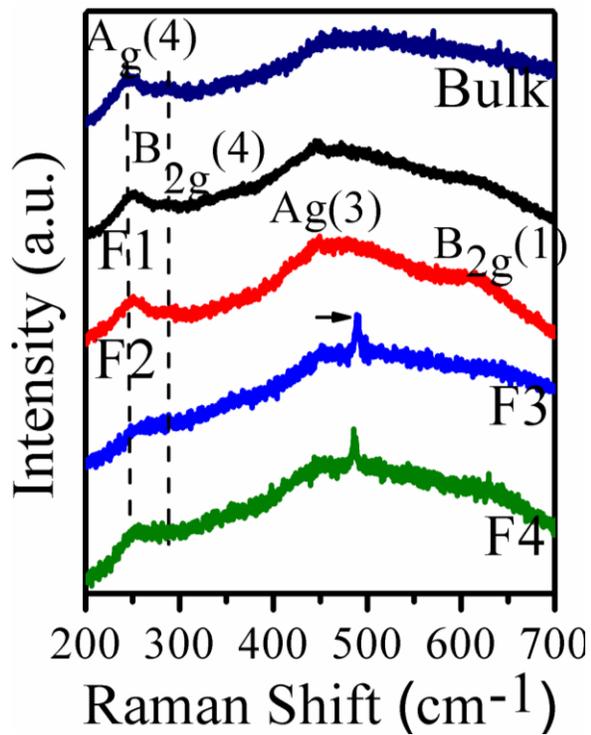

**Figure 5:** Room temperature Raman spectra of F1, F2, F3 and F4 films along with bulk compound in unpolarized geometry using He-Ne (632.8 nm) excitation source.

**Figure 6:**

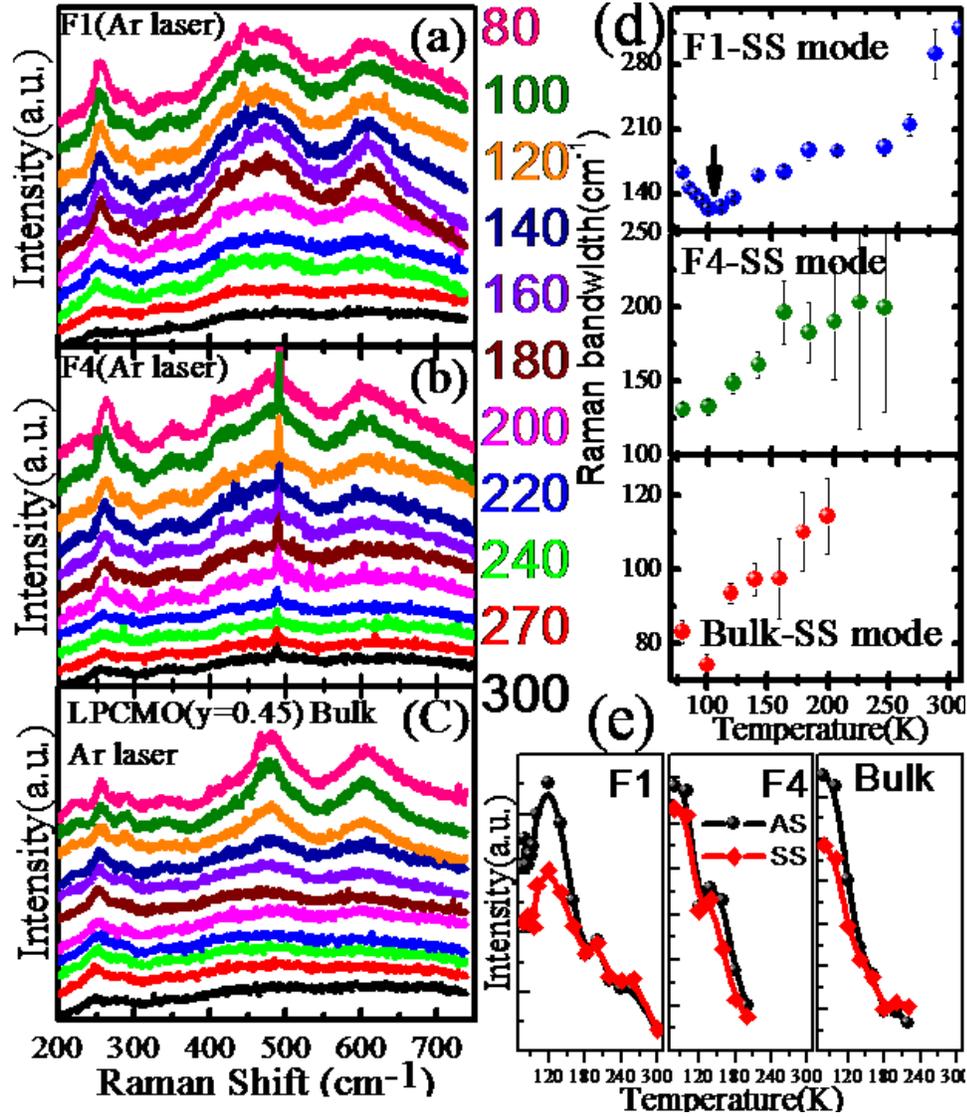

**Figure 6:** (a)-(c) The Low temperature Raman spectra of F1 and F4 films along with bulk compound collected from 300K to 80K using 488 nm excitation source. (d)-(e) Temperature dependence of the linewidth and intensity of the AS and SS modes for F1, F4 and bulk compound. Arrow depicts anomalous behavior of the SS Raman mode width for F1 film below $T_{MI}$.

**Figure 7:**

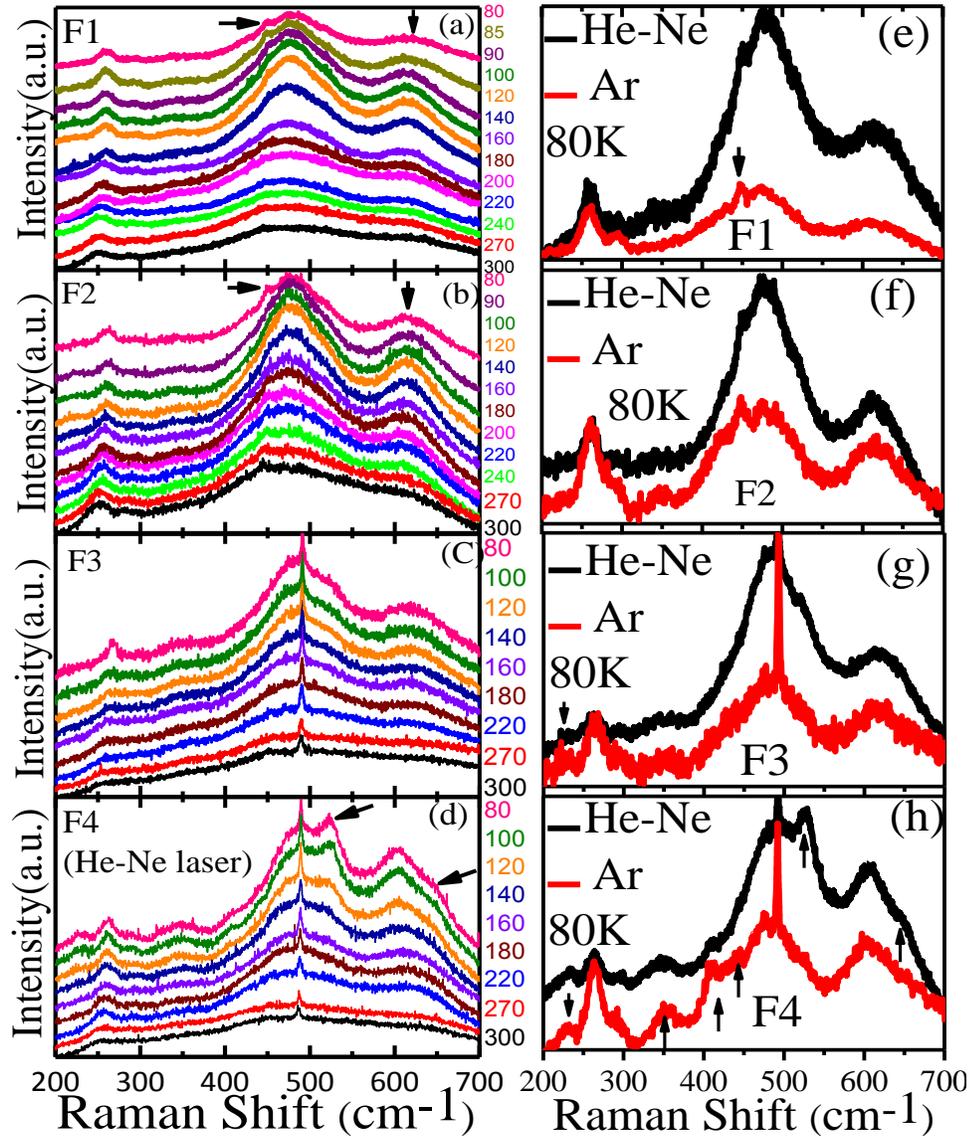

**Figure 7:** (a)-(d) The Low temperature Raman spectra of all the four LPCMO films (F1, F2, F3 and F4) collected from 300K to 80K using 632.8nm excitation source. (e)-(g) Comparison of Raman spectra at 80K for all the four films collected using 632.8nm and 488nm excitation lines. Appearance of new peaks depicted by arrows reveals structural rearrangement below $T_{MI}$ and $T_{CO}$ for F1 and F4 films respectively.

**Table 1 :**
Comparison of T* (2T cooling) and divitrification finish temperature $T_{df}$ for 1T cooled-2T warming curve.

| Film name | $T_{2T}^*$(K) | $T_{df}$(K) |
|---|---|---|
| LPCMO/NGO(200nm)[19] | 129.88 | 44.87 |
| LPCMO/LAO(175nm) | 163.54 | 37.08 |
| LPCMO/LAO(250nm) | 169.57 | 34.73 |